\documentclass[preprint2]{emulateapj}

\newcommand{\HI}{\ion{H}{1}}
\newcommand{\OII}{[\ion{O}{2}]}
\newcommand{\kms}{km\,s$^{-1}$}
\newcommand{\Msun}{M$_\sun$}
\newcommand{\SDSSa}{SDSS~J210258.87+103300.6}
\newcommand{\SDSSb}{SDSS~J230743.41+152558.4}
\newcommand{\SDSSc}{SDSS~J233453.20+145048.7}

\shorttitle{The \HI\ content of E+A galaxies}
\shortauthors{Buyle, Michielsen, De Rijcke, Pisano, Dejonghe, Freeman}

\begin{document}

\title{The \HI\ content of  E+A galaxies}

\author{P. Buyle\altaffilmark{1}, D. Michielsen\altaffilmark{2}, S. De
Rijcke\altaffilmark{1,5},
D.J. Pisano\altaffilmark{3,6}, H. Dejonghe\altaffilmark{1},
K. Freeman\altaffilmark{4}}

\altaffiltext{1}{Sterrenkundig Observatorium, Universiteit Gent,
Krijgslaan 281, S9, B-9000, Ghent, Belgium; Pieter.Buyle@UGent.be,
Sven.DeRijcke@UGent.be, Herwig.Dejonghe@UGent.be}
\altaffiltext{2}{School of Physics and Astronomy, University of
Nottingham, Nottingham NG7 2RD, UK; dolf.michielsen@nottingham.ac.uk}
\altaffiltext{3}{Naval Research Laboratory, USA;
Daniel.Pisano@nrl.navy.mil} \altaffiltext{4}{Research School of
Astronomy and Astrophysics, Australia National University, Cotter Rd,
ACT2611, Australia; kcf@mso.anu.edu.au}
\altaffiltext{5}{Post-doctoral Fellow of the Fund for Scientific Research -
  Flanders, Belgium (F.W.O.)}
\altaffiltext{6}{National Research Council Postdoctoral Fellow}
\begin{abstract}
We present deep single-dish \HI\ observations of a sample of six
nearby E+A galaxies ($0.05<z<0.1$). A non-negligible fraction of a
local sample of E+As are detected in HI. In four galaxies, we have
detected up to a few times $10^9$ \Msun\ of neutral gas. These E+A
galaxies are almost as gas-rich as spiral galaxies with comparable
luminosities.
There appears to exist no direct correlation between the
amount of \HI\ present in an E+A galaxy and its star-formation rate as
traced by radio continuum emission. Moreover, the end of the starburst
does not necessarily require the complete exhaustion of the neutral
gas reservoir. Most likely, an intense burst of star formation
consumed the dense molecular clouds, which are the sites of massive
star formation.  This effectively stops star formation, even though
copious amounts of diffuse neutral gas remain. The remaining \HI\
reservoir may eventually lead to further episodes of star
formation. This may indicate that some E+As are observed in the
inactive phase of the star-formation duty cycle.
\end{abstract}


\section{Introduction}\label{intro}

At $z \approx 0.5$, galaxy clusters contain a population of blue,
distorted galaxies that is missing in local clusters: the so-called
Butcher-Oemler effect \citep{bo78}. In their spectroscopic study of
blue galaxies in three clusters at z$\sim$0.31 \citet{cs87} found that
sixty percent of these blue galaxies (classified as Type 3 galaxies)
have optical spectra characterised by strong Balmer absorption lines,
typical for a very young stellar population, but with weak, if any,
emission lines, such as \OII~$\lambda$3727{\AA}. Since these spectra
are a superposition of an old stellar population, resembling that of
an elliptical 'or E' galaxy, and a young stellar population, dominated
by A stars, these galaxies are called E+A galaxies. In the notation of
\citet{dr99}, these galaxies would be classified as k+a/a+k,
satisfying the criteria EW(\OII)$< 5${\AA} and
EW(H$\delta)>$3{\AA}. While at $z \approx 0.4$, about 20\% of all
cluster galaxies are classified as E+As \citep{belloni95}, they
constitute less than 1\% of the present-day clusterpopulation
\citep{fabricant91}.

Apparently, these galaxies are observed during a quiescent phase,
which explains the lack of emission lines, soon after a vigorous
starburst, which explains the strong Balmer absorption lines. Due to
the short lifetimes of the stars causing the Balmer absorption, the
starburst must have ended no more than $\sim 1$~Gyr ago
\citep{dg83,p99}. E+As often have disturbed morphologies, e.g. tidal tails,
suggestive of recent merger or interaction events. They span the whole
morphological range, from bulge-dominated with underlying disks to
disk-dominated \citep{tran03,yang04}. Their high surface brightness
sets them apart from the elliptical and lenticular galaxies in the
Fundamental Plane. Over time, fading of the stellar population will
drive them towards the locus of the E/S0s \citep{yang04,p99}. Internal
velocity dispersions of galaxies classified as E+As appear to increase
as a function of redshift, going from $\sigma \simeq 150$ \kms\ at $z
= 0.3$ to $\sigma \simeq 250$ \kms\ at $z=0.83$ \citep{tran03}. This
trend suggests that massive galaxies undergo an E+A phase, i.e. are
observed in a post-starburst phase, at earlier cosmic times than less
massive ones. This is reminiscent of the ``down-sizing'' phenomenon in
star-forming galaxies \citep{c96}, according to which the masses of
galaxies hosting star formation decrease as the Universe ages.

\begin{table*}
\begin{center}
\caption{Properties of the E+A sample.}\label{EAproperties}
{\scriptsize \begin{tabular}{lccccccc}
\hline
Galaxy &  RA (J2000)\tablenotemark{a} & $\delta$ (J2000)\tablenotemark{a} & $M_B$ & $v_{\rm hel}$\tablenotemark{b} & $\Delta v$ & $\int S(v)\,dv$\tablenotemark{c}& H\,{\sc i} mass\tablenotemark{b}\\
       & (h,m,s) & ($^\circ$,$\arcmin$,$\arcsec$) & (mag) & (\kms) & (\kms) & (Jy~km~s$^{-1}$) & ($10^9$\, \Msun)\\  
\hline
\SDSSa & 21 02 58.9 & +10 33 01 & -21.7& 27821 & 440 & $0.18 \pm 0.02$ & $6.5 \pm 0.8$ \\
\SDSSb & 23 07 43.4 & +15 25 58 & -20.4& 20894 & 240 & $0.04 \pm 0.01$ & $0.9 \pm 0.3$ \\
\SDSSc & 23 34 53.2 & +14 50 49 & -20.2& 19388 &  380 & $0.15 \pm 0.02$ & $2.7 \pm 0.3$c\\
LCRS B101120.1-024053 (EA17) & 10 13 52.4 & -02 55 48 & -18.2 & 18258 & & $<0.18$ & $< 2.9 $ \\
LCRS B002018.8-415015 (EA18) & 00 22 47.1 & -41 33 37 & -18.9 & 17941 & 660 & $0.15 \pm 0.02$ & $2.3 \pm 0.3$ \\
LCRS B020551.6-453502 (EA19) & 02 07 49.7 & -45 20 50 & -18.9 & 19186 &  & $<0.07$ & $< 1.2 $ \\
\hline
\hline
\end{tabular}\vspace*{1cm}\\
\begin{tabular}{lccccc}
\hline
Galaxy &  [O{\sc ii}]\tablenotemark{d} & H$\delta$\tablenotemark{d} & $r_e$ & S\'ersic index & SFR\\
       &({\AA}) & ({\AA}) & (kpc) &  & (\Msun/yr)\\  
\hline
\SDSSa & -0.68& 5.13 & 4.6 & 3.2  & $10.2^{^{+15.7}_{-6.5}}$\\
\SDSSb & -0.92& 6.78 & 2.0 & 3.1 & $1.2^{^{+1.8}_{-0.8}}$ \\
\SDSSc & -1.27& 5.18 & 2.1 & 3.3 & $5.6^{^{+10.1}_{-3.7}}$ \\
LCRS B101120.1-024053 (EA17) & 1.68 & 6.92 & 1.5 & 2.7/1.3  &  $< 8.2^{^{+18.1}_{-5.9}}$\\
LCRS B002018.8-415015 (EA18) & 1.75 & 5.96 & 1.7 & 2.2  &  $5.3^{^{+10.4}_{-3.7}}$\\
LCRS B020551.6-453502 (EA19) & 0.98 & 6.08 & 2.1 & 1.1  & $<1.8^{^{+2.7}_{-1.1}}$\\
\hline
\hline
\end{tabular}}
\tablenotetext{a}{The coordinates are taken from Goto et al. (2003) in
case of the SDSS sample and from Zabludoff et al. (1996) in case of
the LCRS sample.}  \tablenotetext{b}{The distance $D$, required for
calculating the H{\sc i} mass, is estimated as the Hubble distance, $D
= v_{\rm hel}/H_0$, with $H_0 = 70$ \kms\,Mpc$^{-1}$.}
\tablenotetext{c}{For the non-detected galaxies, EA17 and EA19, we
used a velocity width of 450~km~s$^{-1}$.} \tablenotetext{d}{[O{\sc
ii}] and H$\delta$ equivalent widths ($\langle {\rm
H}\beta\gamma\delta\rangle$ for the LCRS sample); negative values for
{\OII} indicate absorption}
\end{center}
\end{table*}
Although the majority of low-$z$ E+A galaxies have a smooth light
distribution, many of them also show slightly disturbed morphologies
(e.g. warps and dust lanes). Based on this, \citet{zab96} argued that
the E+A phase is the aftermath of a vigorous starburst, triggered by a
major merger or interaction.  This is corroborated by a study of E+A
galaxies drawn from the 2dFGRS catalogue \citep{blake04}. About three
quarters of all E+A galaxies are found in the field (i.e. outside
clusters), simply because most of the galaxies in the universe do not
reside in clusters. However, the \emph{fraction} of E+As is four times
higher in clusters than in the field \citep{zab96,tran04}. The spatial
distribution of E+A galaxies in clusters is more extended than that of
quiescent galaxies, but less extended than that of emission-line
galaxies \citep{dr99}, suggesting that processes such as galaxy
harassment or ram-pressure stripping, which are specific to clusters,
can also cause the E+A phenomenon. Based on the 2dFGRS E+A sample,
\citet{blake04} find that the distribution of {\em local} environments
of E+A galaxies closely traces that of the ensemble of 2dFGRS galaxies
and conclude that whatever causes the E+A phenomenon, must be a very
local mechanism, such as encounters of galaxy pairs. This is
corroborated by a recent analysis of the environments of E+As selected
from the SDSS \citep{goto05b}.

Most E+As have E/S0-like morphologies, with a small fraction of
ongoing interactions. Their luminosity distribution is more similar to
the distribution of spectroscopically defined elliptical galaxies than
to the luminosity distribution of the ensemble of 2dFGRS galaxies
\citep{blake04}. However, not all E+As can be associated with mergers
and, obviously, more than one evolutionary pathway can lead to a
post-starburst galaxy \citep{tran03,dr99}. Numerical simulations show
that E+As can indeed be formed via a major merger of two gas-rich
spiral galaxies \citep{bekki05}. A disk-disk merger event then
triggers a starburst, which consumes, or, by feedback, expels most of
the available gas and then subsides. The young stars then dominate the
optical spectrum for the following few hundred Myr while emission
lines are absent. During this time-span, a galaxy would be classified
as an E+A. In this case, one expects star-formation to be centrally
concentrated, leading to radially decreasing Balmer line strengths
\citep{pracy05}. Alternatively, star formation can be truncated more
or less instantaneously over the whole disk of the galaxy without a
starburst, e.g. by the gas being swept away by ram pressure
stripping. In this case, as the young star population fades, the older
bulge population causes the strengths of the Balmer lines to be
radially increasing \citep{pracy05}.

The red colours of some H$\delta$-strong E+As cannot be explained by
any plausible starburst model \citep{cs87,blake04}, leaving heavy dust
extinction as the only viable explanation. This hypothesis can be
tested by using dust insensitive tracers of star formation. Since
radio continuum emission is synchrotron radiation from electrons
accelerated in supernova remnants, it is an indirect tracer of star
formation. \citet{mo01} observed part of the \citet{zab96} sample and
detected radio-continuum emission in only two out of fifteen
galaxies. \citet{smg99} detect five out of eight post-starburst
galaxies at radio wavelengths. Radio-continuum observations of a
sample of 36 E+As drawn from the SDSS yielded no detections
\citep{goto05}. 
If this apparent lack of ongoing star-formation in E+As is true, then
no dust obscuration, hiding the star-formation sites, needs to be
invoked. Near-infrared studies \citep{gal00,bal05} have shown that the
u$-$g and r$-$k colours, and the H$\delta$ line-strengths of E+As can
be well explained by dust-free models in which more than 5\% of the
stellar mass has recently been produced in a starburst. Hence, the
presence of dust is still uncertain because of these contradictory
observations.

Up to now only one search for \HI\ in E+As was conducted
\citep{chang01}. VLA observations of five E+As from the sample of
\citet{zab96} resulted in the detection of only one field E+A galaxy,
EA1, with a total H{\sc i} mass of $7.1 \pm 0.4 \times 10^9\,M_\odot$
(assuming $H_0 = 70$ \kms\,Mpc$^{-1}$). For the four other galaxies, upper
limits of order $10^9\,M_\odot$ could be derived. EA1 consists of two
components, and is most likely a merger remnant. However, other
galaxies in this sample are also optical mergers but they do not
contain detectable amounts of gas.  We started an \HI\ study of a
sample of E+A galaxies in order to constrain the amount of neutral gas
present in these systems. In section~\ref{obs} we describe the sample
and the observations. The results are presented in
section~\ref{results}. We discuss these results in
section~\ref{discussion} and summarise our conclusions in
section~\ref{concl}.

\begin{figure*}
\begin{center}
\includegraphics[scale=0.6]{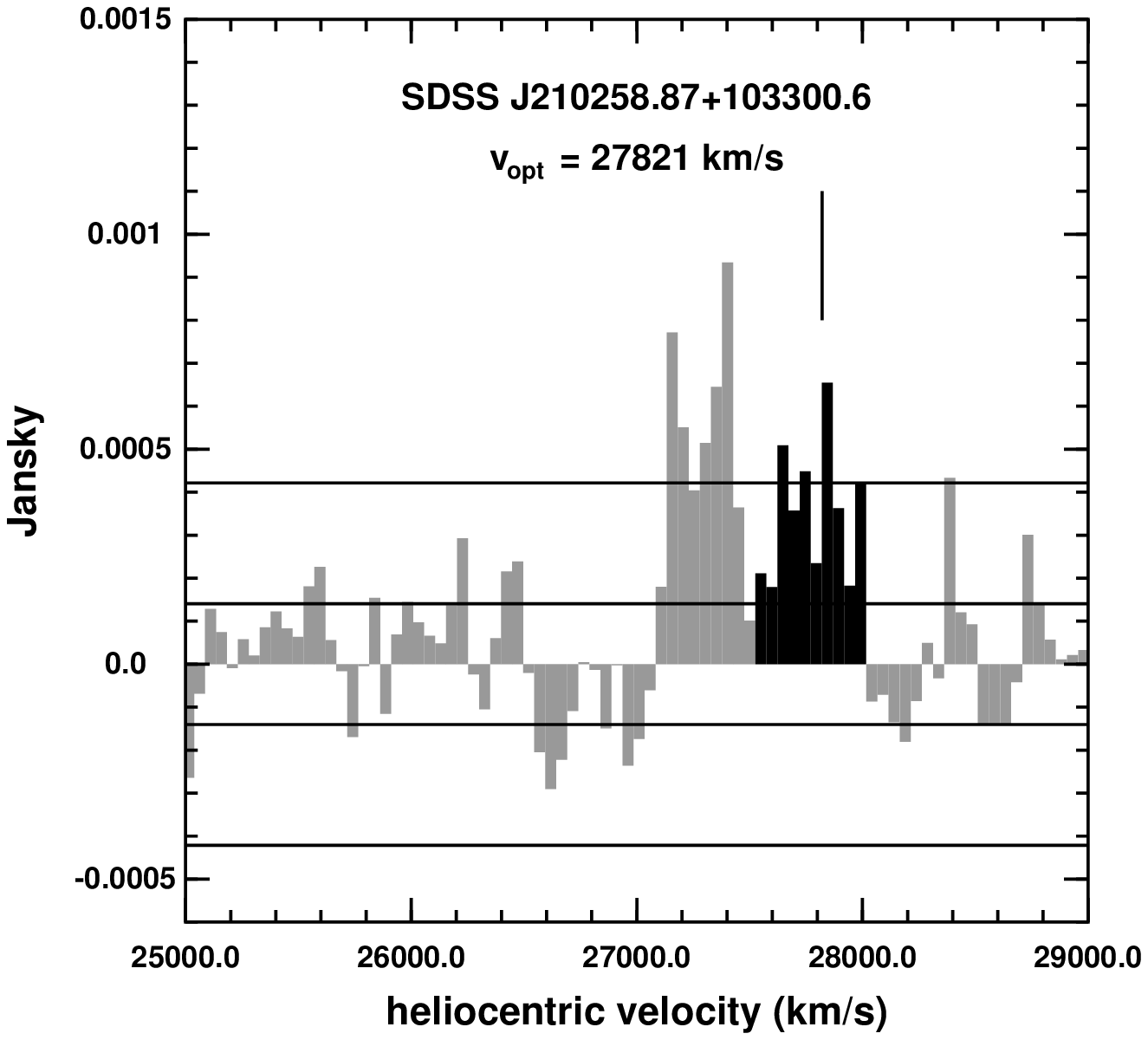}
\includegraphics[scale=0.6]{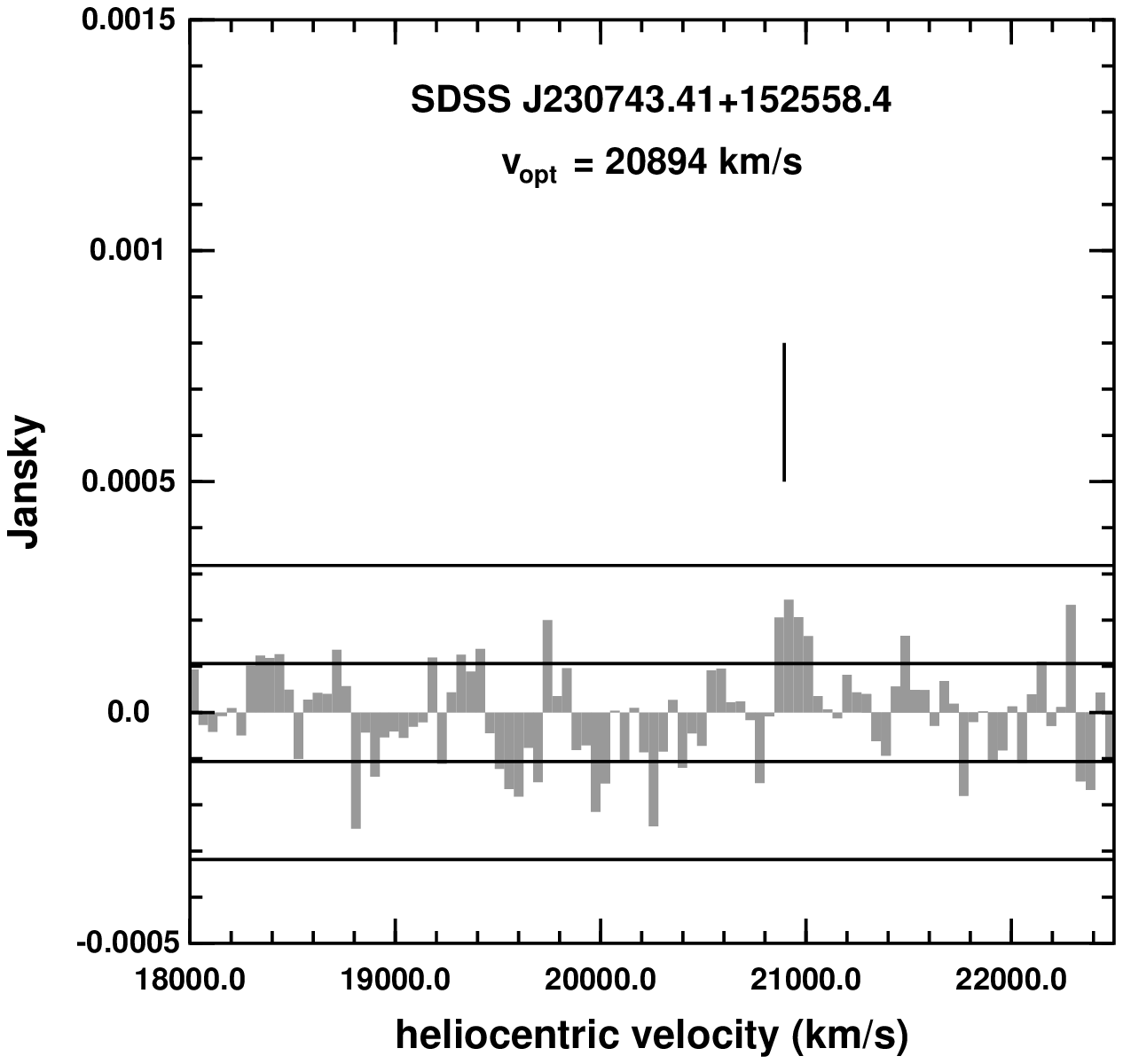} \\
\includegraphics[scale=0.6]{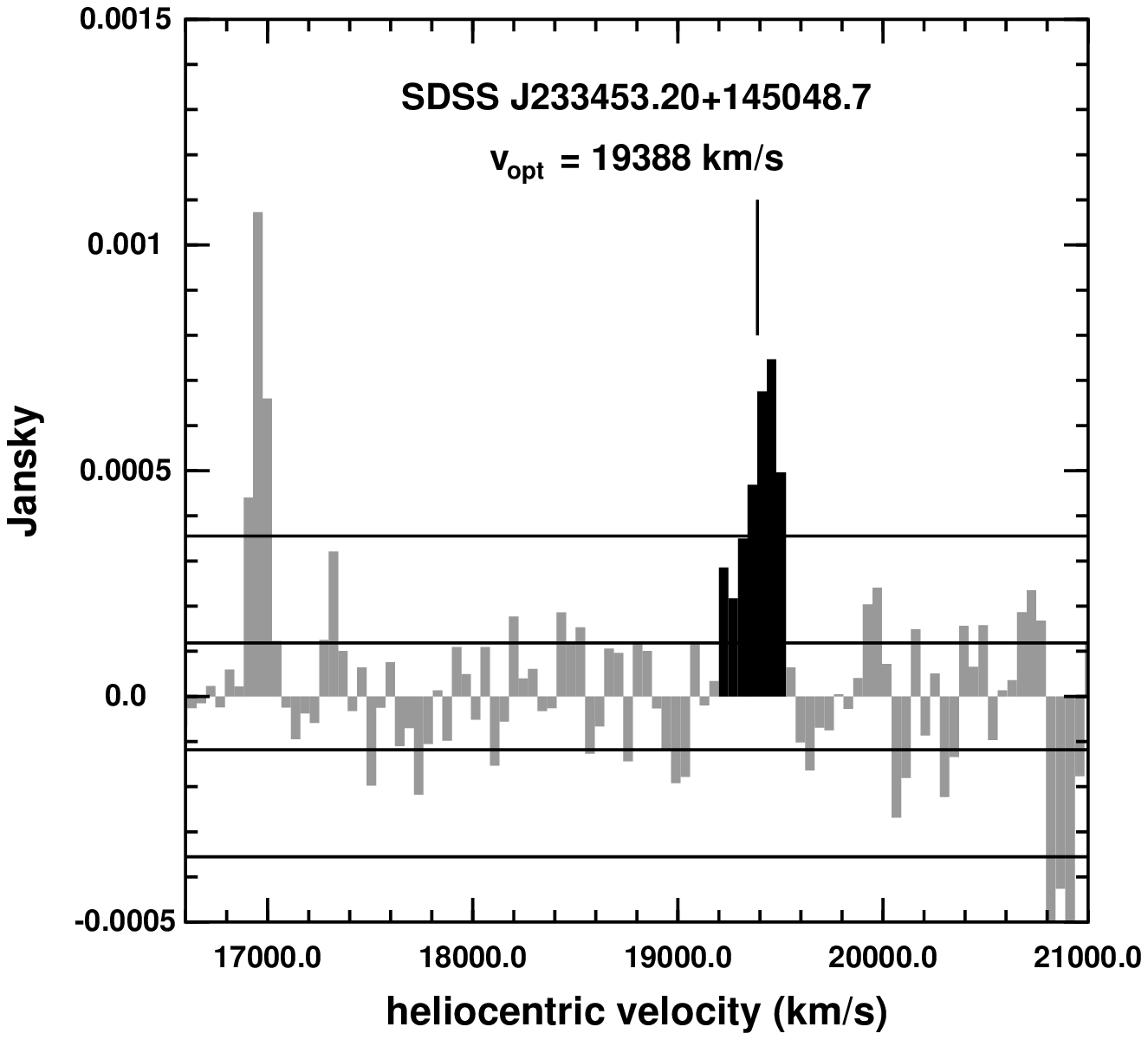}
\includegraphics[scale=0.6]{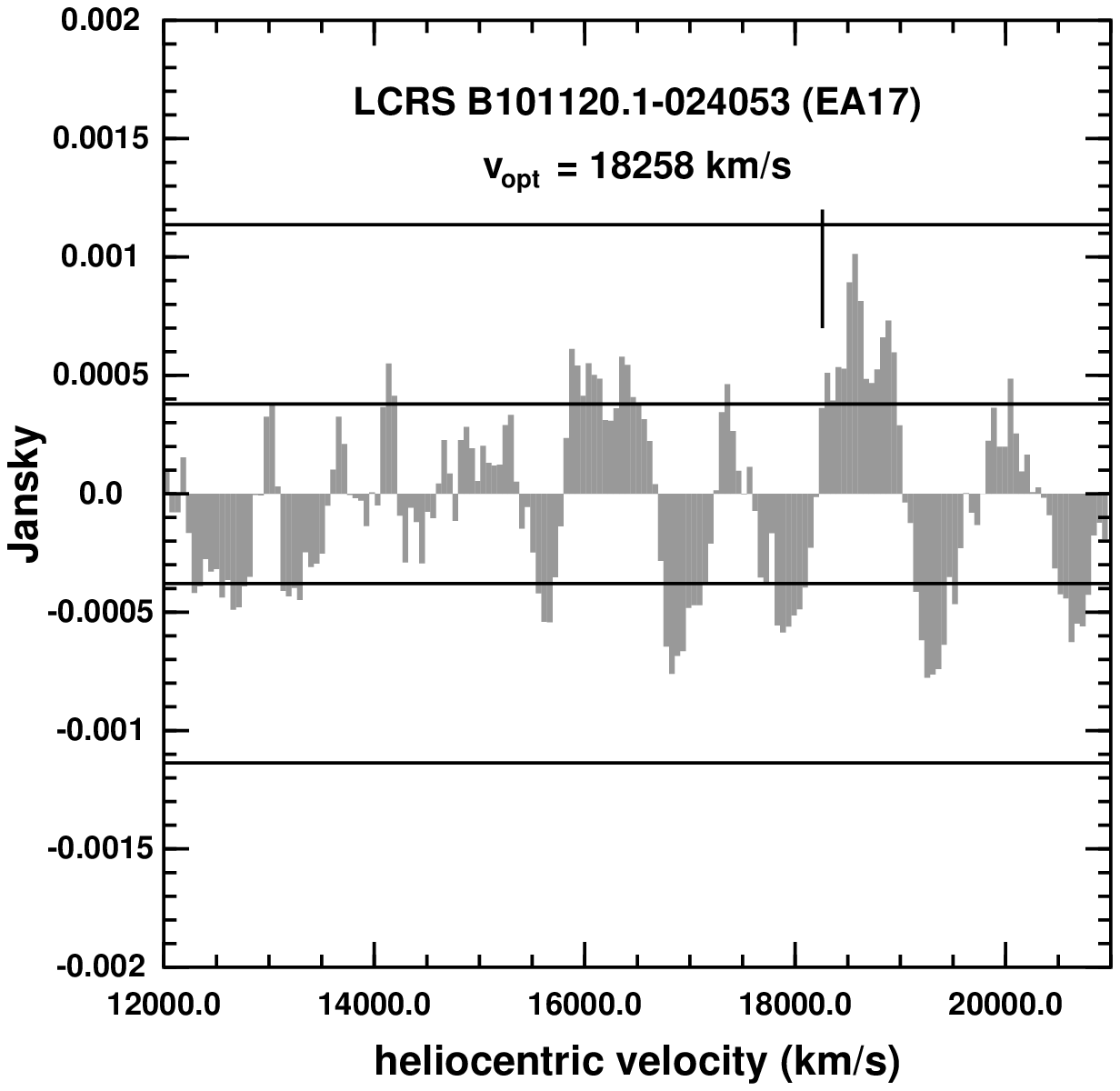} \\
\includegraphics[scale=0.6]{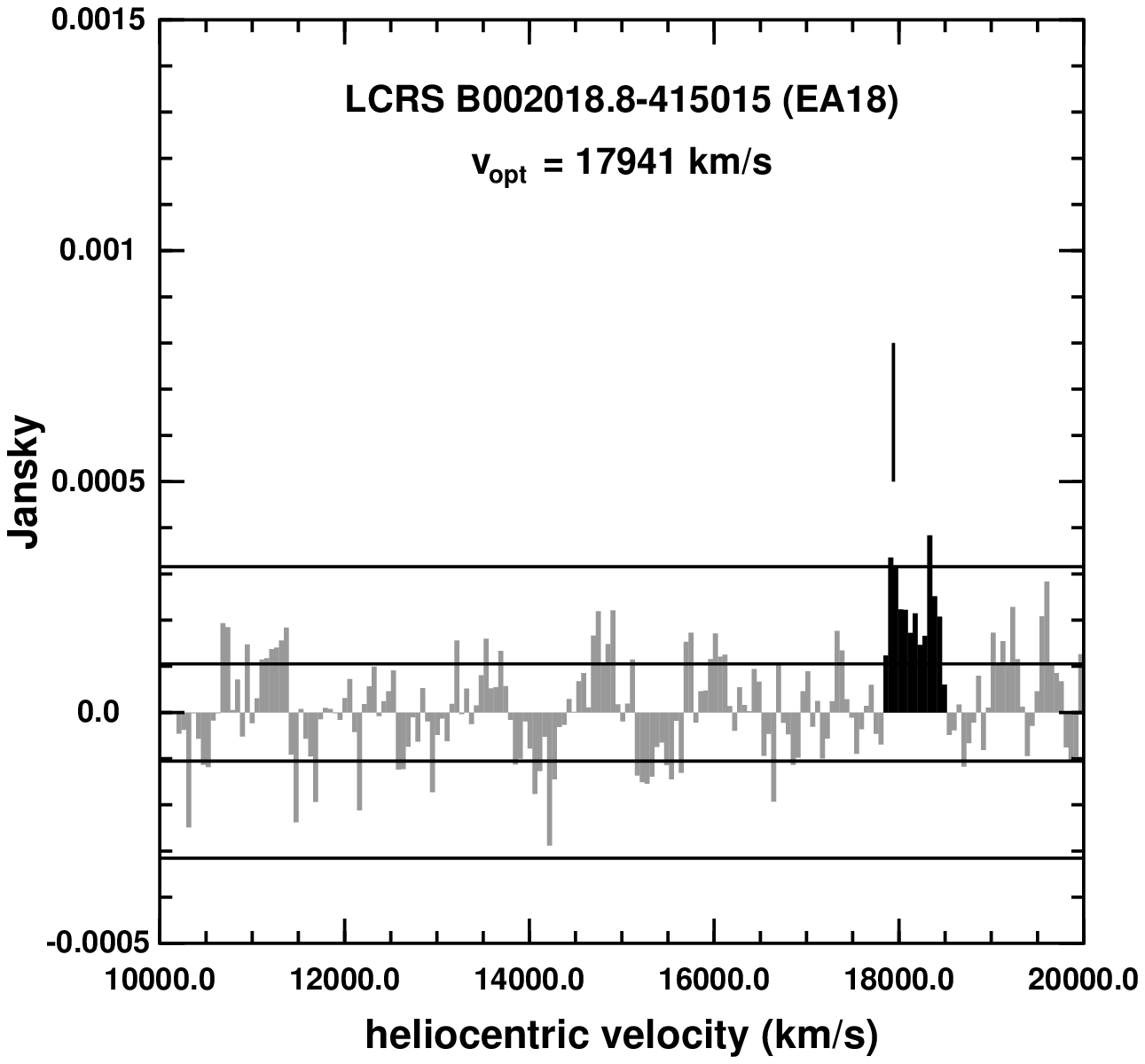}
\includegraphics[scale=0.6]{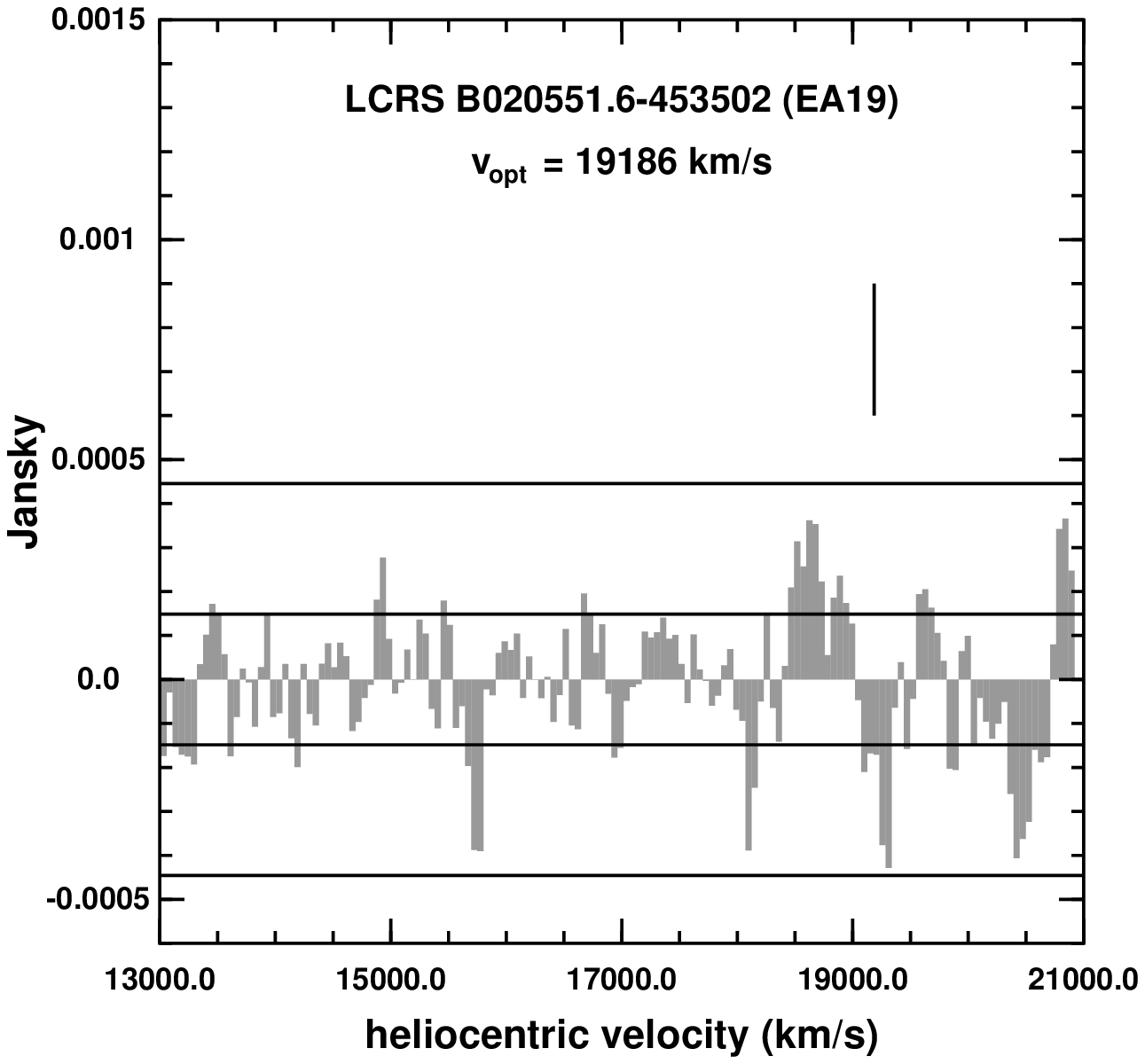}
\caption{The \HI\ spectra of our E+As. The galaxies observed with
Arecibo, SDSS~J210258.87+103300.6, SDSS~J2334543.20+145048.7,
SDSS~J2330743.41+152558.4, have been rebinned to a velocity resolution
of 50 \kms. The Parkes dataset, LCRS~B101120.1-024053 (EA17),
LCRS~B002018.8-415015 (EA18), LCRS~B020551.6-453502 (EA19), has been
rebinned to a velocity resolution of 53 \kms. The name of the galaxy
is indicated in each panel along with its optical systemic
velocity. The horizontal black lines indicate the -3, -1, 1 and 3
sigma rms noise levels. The vertical black line indicates the optical
velocity as found in Goto et al. (2003) and Zabludoff et
al. (1996). For the three galaxies with clear detections, the bins
corresponding to the galaxy are indicated in black. There is another
galaxy within the beam of SDSS~J210258.87+103300.6 (the double-horn
profile around 27300 \kms).}
\end{center}
\label{HIspectra}
\end{figure*}

\section{Observations}\label{obs}

\subsection{The sample}

Since {\em (i)} the occurrence of the E+A phenomenon seems to be
determined predominantly by the local environment, {\em (ii)} the
mechanisms triggering the E+A phenomenon, i.e. mergers, interactions,
ram-pressure stripping, act both at high and low redshift, and {\em
(iii)} for a given flux, the {\HI} mass scales with distance squared,
we opted to use distance as our main selection criterion to maximise
our chances of a detection. We selected the three closest E+A galaxies
from the large catalogue compiled by \citet{goto03} from the SDSS, and
the three closest E+A galaxies from the compilation of
\citet{zab96}. The E+A sample selected from the SDSS by \citet{goto03}
satisfies very strict criteria and contains only galaxies with
EW(H$\delta)>4{\rm\AA}+\Delta{\rm EW}({\rm H}\delta)$, with
$\Delta{\rm EW}({\rm H}\delta)$ the 1$\sigma$ error on the H$\delta$
equivalent width, and no detectable \OII\ and H$\alpha$ emission,
quantified by the constraints EW(\OII$)<\Delta$EW(\OII) and
EW(H$\alpha)<\Delta{\rm EW}({\rm H}\alpha)$, respectively. The
Zabludoff sample satisfies the following
criteria:~EW(H$\beta\gamma\delta)>5.5{\rm\AA}$ and
EW(\OII$)<2.5${\AA}, with EW(H$\beta\gamma\delta$) the mean
equivalent width of the H$\beta$, H$\gamma$, and H$\delta$ absorption
lines. Table~\ref{EAproperties} summarises the properties of the
galaxies in our sample.

\subsection{Arecibo observations}\label{arecibo}

We observed the galaxies SDSS~J210258.87+103300.6, {\SDSSb}, and
{\SDSSc} for 7.5~h each, including overhead, with the 305m Arecibo
Radio Telescope\footnote{The Arecibo Observatory is part of the
National Astronomy and Ionosphere Center, which is operated by Cornell
University under a cooperative agreement with the National Science
Foundation.} in Puerto Rico. The observations were scheduled on the
nights of 23$-$25 June, 13$-$15 July and 28$-$30 July 2005. Each
galaxy was observed for 2.5 hours per day during night-time to
minimise solar interference. We used the L-wide receiver which has an
average system temperature of $\approx 27$~K (depending on the
elevation of the source). We selected the interim correlator in both
linear polarisations to process the data. This resulted in final {\HI}
spectra with a total bandwidth of 25~MHz and 12.5~MHz divided over
1024 channels, resulting in a velocity resolution of 6.3 {\kms} and
3.15 {\kms} respectively. For {\SDSSb} and {\SDSSc}, we used the radar
blanker to minimise the effect of the FAA Airport radar at 1330~MHz
and 1350~MHz. The beam size of the L-wide receiver is
$3.1'\times3.5'$. We applied the standard position-switching
algorithm. Each galaxy was observed for 5 minutes, followed by a 5
minute offset by 5$'$ in right ascension to blank sky, such that we
tracked the same azimuth and zenith angle as the on-source scan. This
mode was used for all galaxies. We reduced the data by means of the
standard Arecibo {\tt IDL} routines, written by P. Perrilat. We
calibrated each bandwidth individually and the polarisations were
averaged together before a second-order baseline was fit across the
interference-free part of the spectrum. We checked, with NED, the
recession velocities of all galaxies inside the Arecibo beam to avoid
confusion with other objects. The results are presented in
section~\ref{results}.

\subsection{Parkes observations}\label{parkes}

We observed the galaxies LCRS~B101120.1-024053 (EA17),
LCRS~B002018.8-415015 (EA18), and LCRS~B020551.6-453502 (EA19) with
the 64m ATNF Parkes Radio Telescope\footnote{The Parkes telescope is
part of the Australia Telescope which is funded by the Commonwealth of
Australia for operation as a National Facility managed by CSIRO.} in
Australia from dusk till dawn from 11 till 15 October 2005. EA17 was
observed during sunrise. We used the Multibeam Correlator in the MB13
configuration (beam-switching mode) which enabled us to observe with 7
beams simultaneously, with one beam on the source while the other six
were pointed adjacently on the sky. The beams were switched in
position each 5 minutes. This way, we derived {\HI} spectra with a
bandwidth of 64~MHz divided over 1024 channels which yields a spectral
resolution of 13.19 {\kms} and a beam width of 14.1$'$. The
integration times, including overhead, were 10~h for EA17, 21~h for
EA18, and 10~h for EA19. This resulted in one clear
3$\sigma$-detection of EA18. All observational quantities are listed
in table~\ref{EAproperties}. The data were reduced by means of the
{\tt Livedata} data reduction pipeline, which is especially developed
for the Multibeam Correlator. No obvious radio interference could be
observed and a second order polynomial was used to fit the spectral
baseline, after masking out a region around the optical
velocity. Afterwards, all data were combined with the help of the {\tt
Gridzilla} software package using the median of the weighted values as
an estimator. Finally, residual baseline ripples were removed by using
the {\tt mbspect} fitting algorithm in the {\tt MIRIAD}
\citep{sault95} software package. Again we checked the recession
velocities of all other objects within the Parkes beam to avoid
confusion. The results are presented in section~\ref{results}.

\begin{figure*}
\begin{center}
\includegraphics[width=7cm]{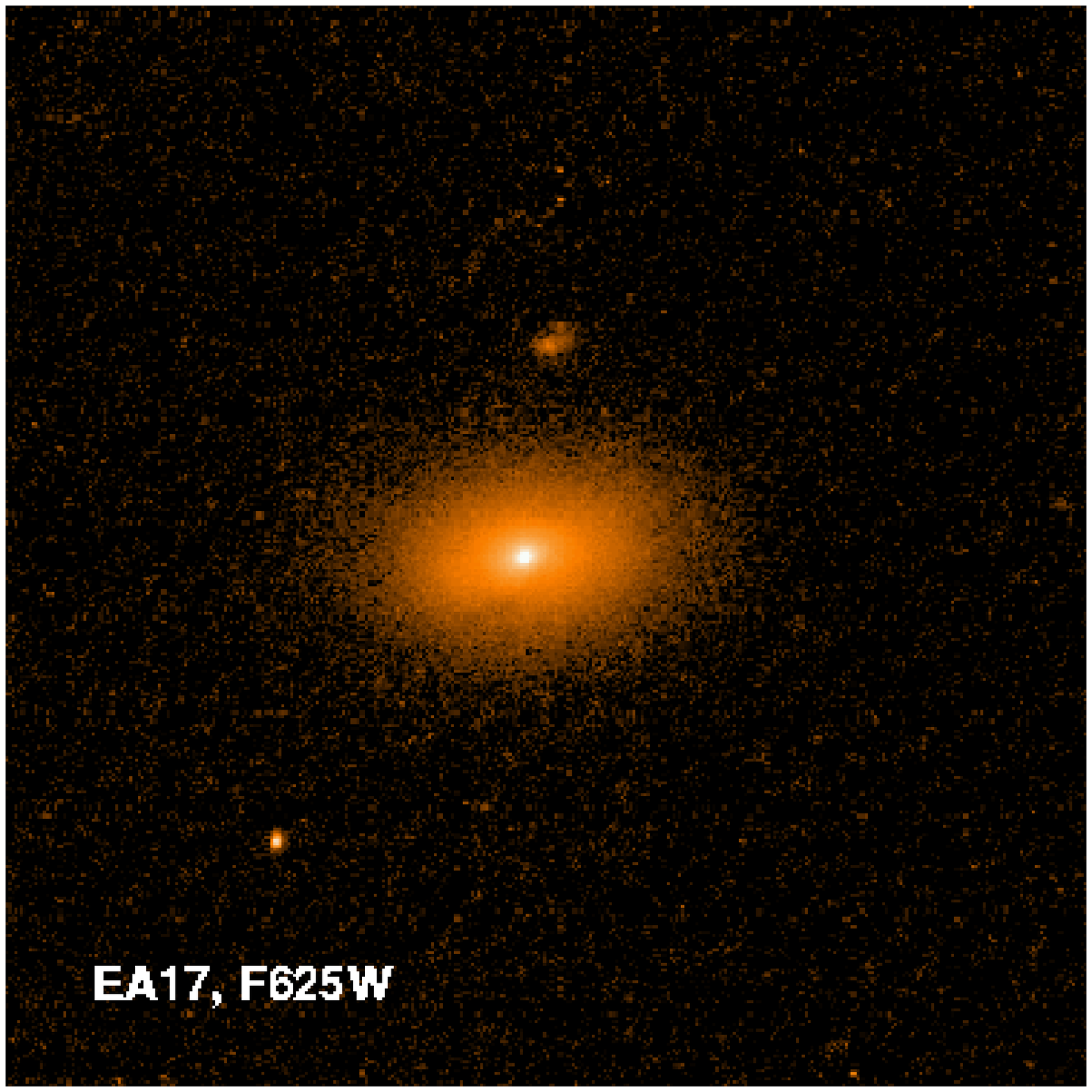}
\includegraphics[width=7cm]{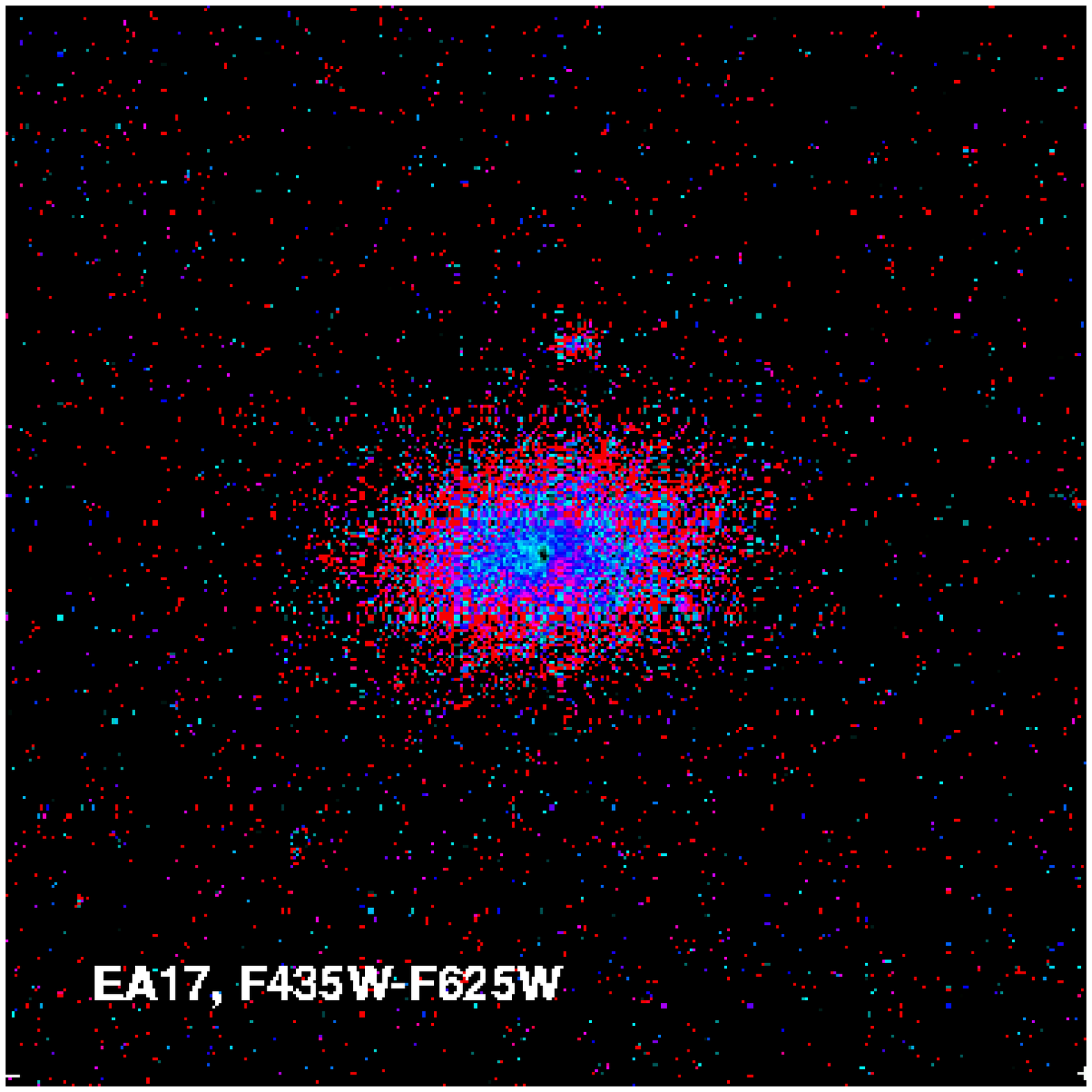} \\
\includegraphics[width=7cm]{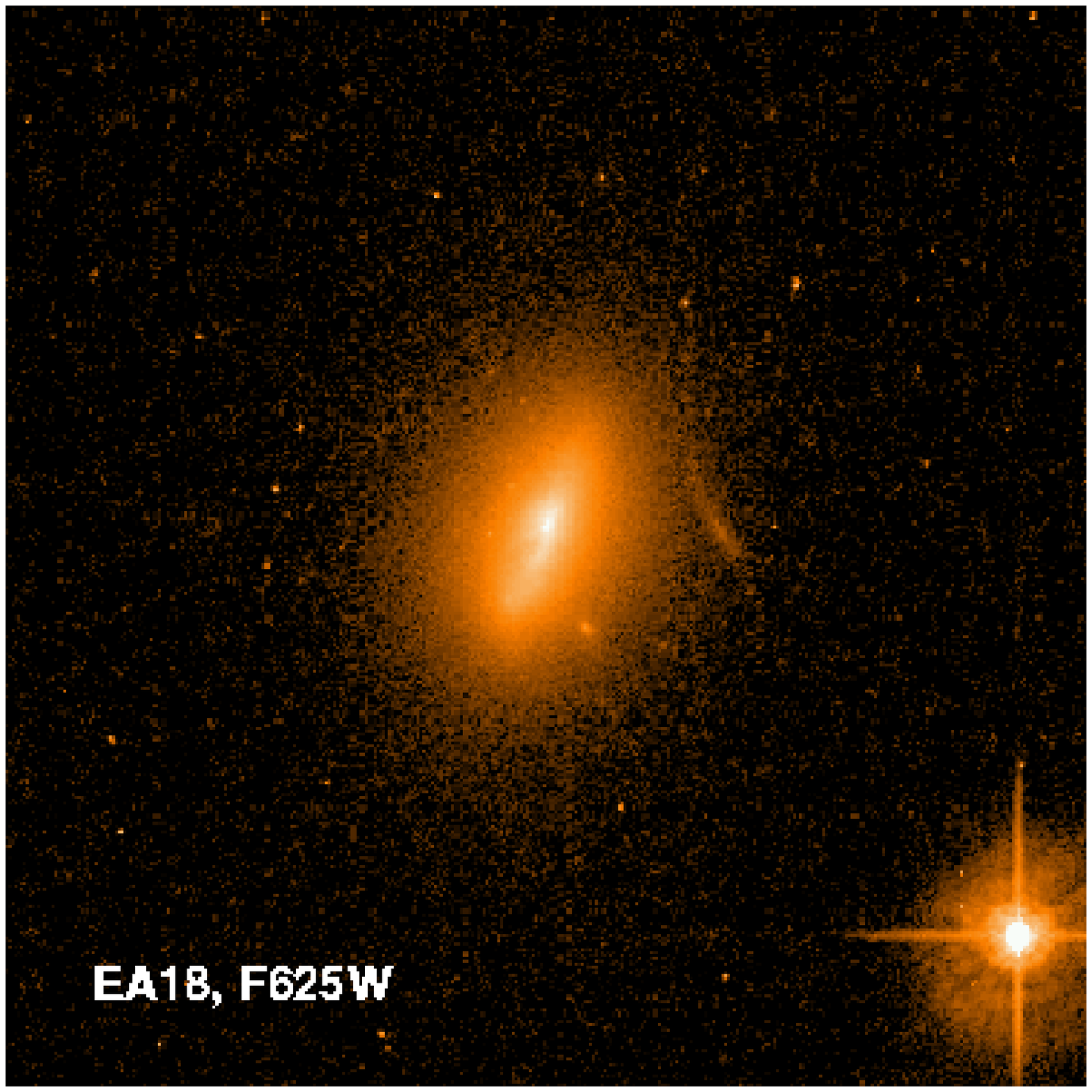}
\includegraphics[width=7cm]{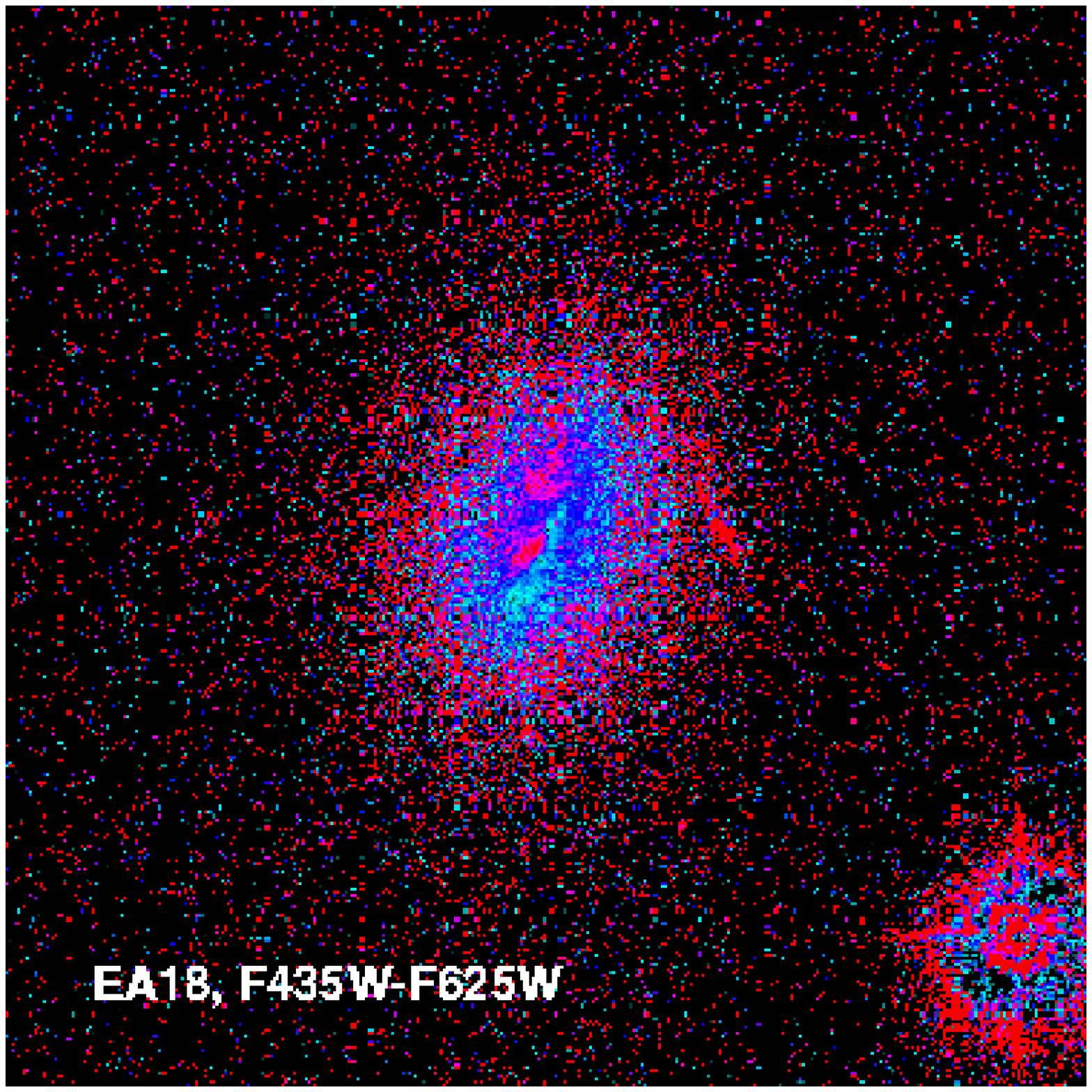} \\
\includegraphics[width=7cm]{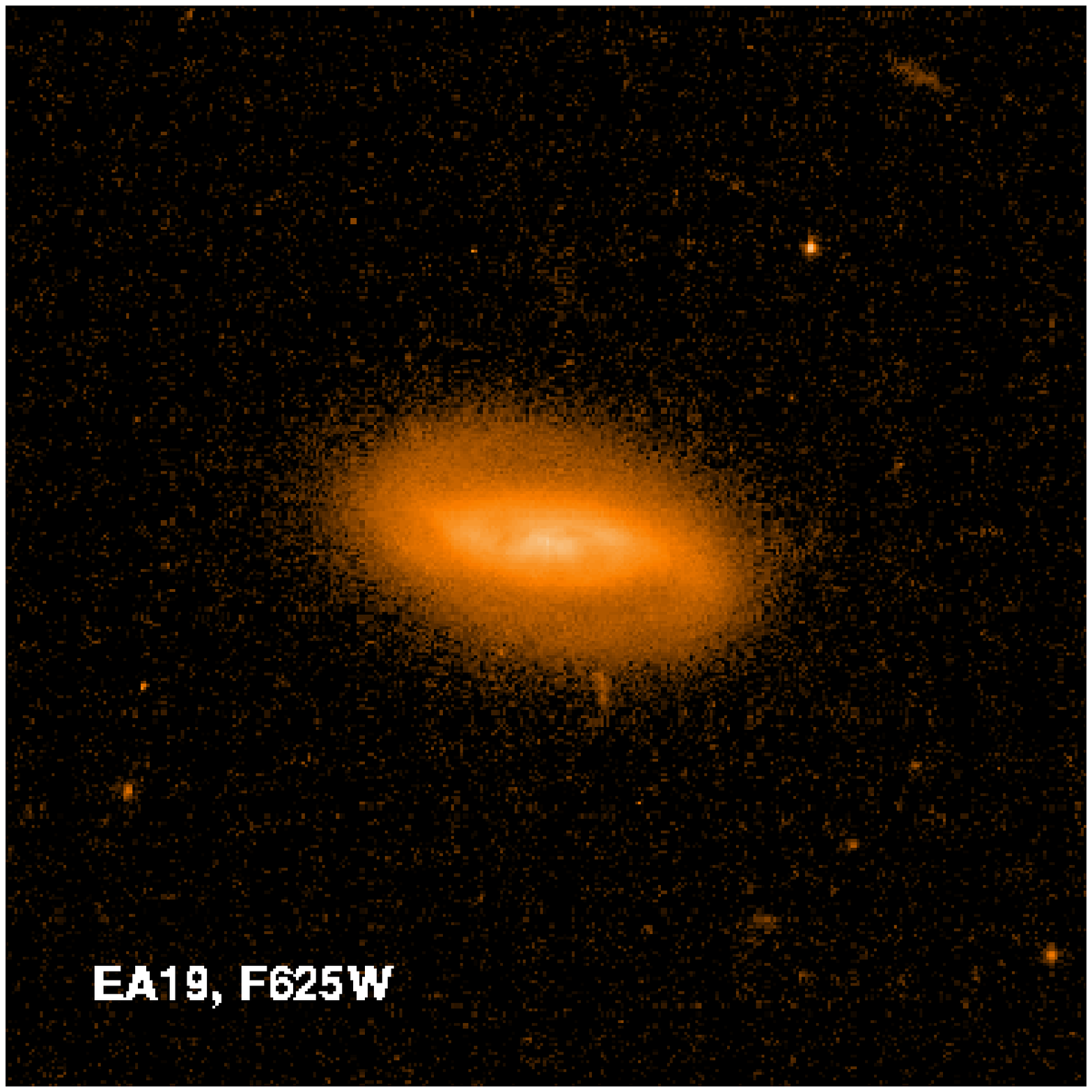}
\includegraphics[width=7cm]{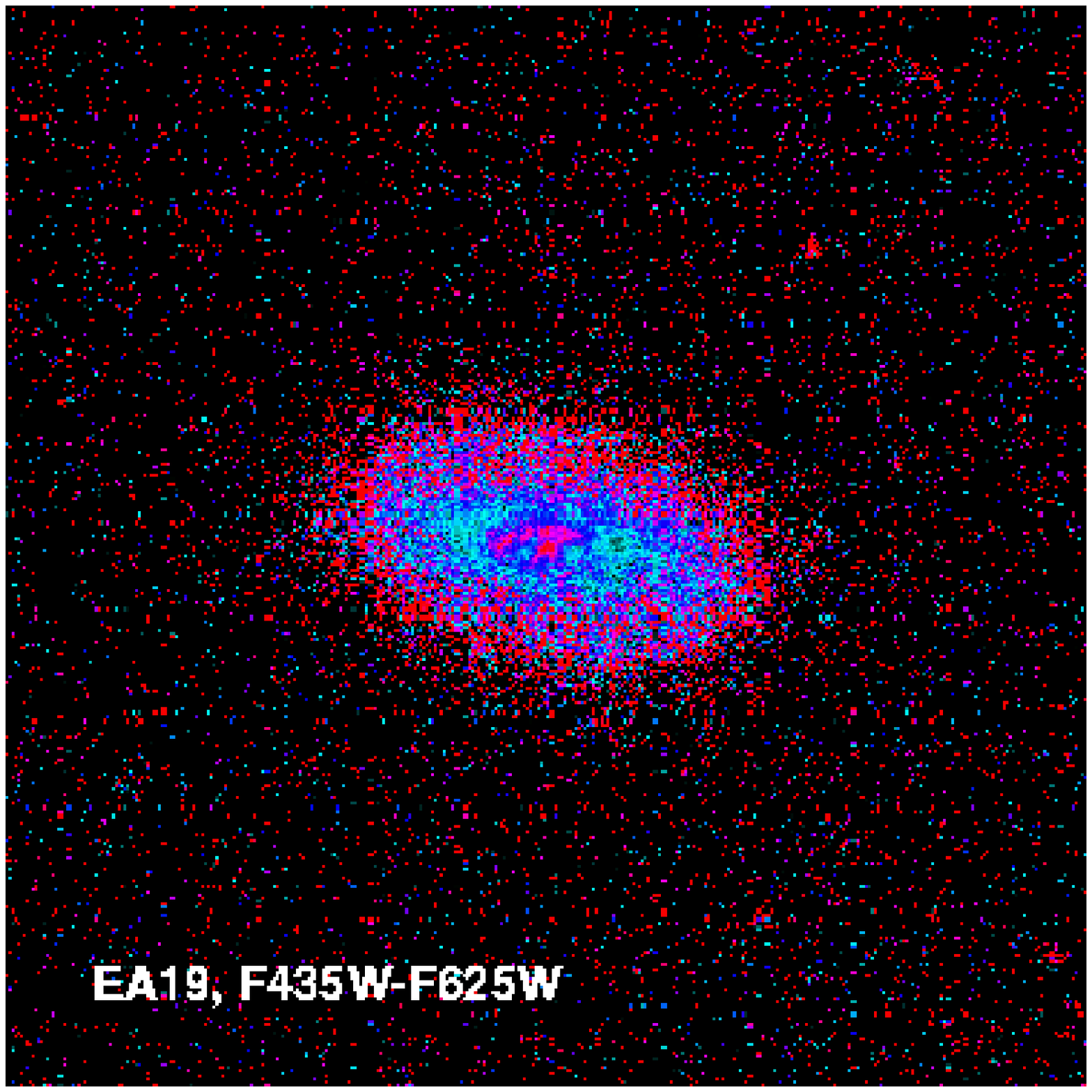}
\end{center}





\caption{Left column:~EA17, EA18, and EA19, imaged with ACS using the
F625W filter. All galaxies are plotted on the same logarithmic colour
scale so their surface brightnesses can be compared directly. Right
column:~F435W$-$F625W colour maps of EA17, EA18, and EA19. All galaxies
show clear evidence for dust patches and lanes. All galaxies are
plotted on the same colour scale, ranging from blue
(F435W$-$F625W$=0.75$~mag) to red (F435W$-$F625W$=1.75$~mag). The
images measure 25$''$ on a side.
\label{ima_EA}}
\end{figure*}

\section{Results}\label{results}

\subsection{\HI\ masses}

In Figure \ref{HIspectra}, we show the \HI\ spectra of our six
E+As. Three of them have a clear 3$\sigma$ detection: SDSS
J210258.87+103300.6, {\SDSSc}, and EA18. In the case of \SDSSb, there
is a peak at the correct velocity, over a velocity width of 240 \kms,
which could be a tentative 2.5$\sigma$ detection of this galaxy. We
did not detect EA17 and EA19. The observations of EA17 were badly
affected by solar interference though.

We calculate \HI\ masses for the detected objects and 3$\sigma$ upper
limits for the undetected galaxies using the formula
\begin{equation}
 M_{\rm H\,I} = 2.36 \times 10^5\,M_\odot\, D^2 \int S(v)\, dv
\end{equation}
with the distance $D$ in Mpc and $\int S(v) \,dv$ the total flux
density in Jy \kms. The distance $D$ was calculated as the Hubble
distance $D = v/H_0$, using $H_0 = 70$ \kms\,Mpc$^{-1}$. We find {\HI}
masses of $6.5 \pm 0.8 \times 10^9\,M_\odot$ for {\SDSSa}, $0.9 \pm
0.3 \times 10^9\,M_\odot$ for {\SDSSb}, $2.7 \pm 0.3 \times
10^9\,M_\odot$ for {\SDSSc}, and $2.3 \pm 0.3 \times 10^9\,M_\odot$
for EA18, as listed in Table~\ref{EAproperties}. To estimate the error
on the {\HI} mass of a galaxy, we generated 50000 statistically
equivalent renditions of its radio spectrum by adding Gaussian noise
to the original spectrum, using the measured 1$\sigma$ noise on the
original datapoints. The mass error is taken to be the rms of the
50000 masses measured from these spectra. The $3\sigma$ upperlimits
for the gas content of EA17 and EA19 are $2.9 \times 10^9\,M_\odot$
and $1.2 \times 10^9\,M_\odot$ respectively, assuming a total velocity
width of 450 {\kms} (the average velocity width of the detected
galaxies).

\subsection{Dust content}

Given the small angular sizes of these galaxies, their dust content
can only be studied accurately with the supreme spatial resolution of
the Hubble Space Telescope (HST). The HST archive contains images of
EA17, EA18, and EA19, obtained with the Advanced Camera for Surveys
(ACS) through the F435W and F625W filters. In Fig.~\ref{ima_EA}, we
show the F625W images and F425W$-$F625W color maps of these
galaxies. From the images it is obvious that the E+A class of galaxies
comprises very different morphological types.

EA17 has a smooth appearance, but it also contains a dust lane, which
is clearly visible in both the colour-map and plain image. The only
galaxy of this sample of three, that was detected in \HI, is EA18,
which is an irregular-looking galaxy with lots of red and blue patches
along the major axis, pointing to dust and/or young stars. EA18
appears also to be warped. Finally, EA19 is a spiral (Sab) galaxy,
with two spiral arms nicely visible in both images. The spiral arms
are slightly bluer than the rest of the disk and the centre contains
blue and red patches, pointing again to dust and/or young stars. 

From just this small sample, it is already clear that dust is present
near the centres of some E+A galaxies. Along with the presence of
neutral {\HI} gas this might indicate that on-going star formation could
in fact be hidden by dust.

\subsection{Environment}

Although E+A galaxies are predominately located in the field
(i.e. outside clusters), their fraction is four times higher in
clusters \citep{zab96,tran04}. By investigating their spatial
distribution inside clusters, \citet{dr99} showed that an E+A
phenomenon can be caused by environmental processes such as galaxy
harassment or ram-pressure stripping. Such processes act mostly on the
neutral hydrogen content of galaxies and can therefore be investigated
by means of our observations.

We adopted the determination of cluster membership of \citet{zab96}
and also checked the number of neighbours within a radius of ~0.5Mpc of
each galaxy. Those E+As that were listed in \citet{zab96} as being
cluster members (EA4 and EA11) have more than 10 known
neighbours. Others, such as EA1, EA3, EA18 have 1 or 2 known neighbours
and are clearly field galaxies, while the remaining ones have 3-7
known neighbours. For the SDSS galaxies from \citet{goto03}, cluster
membership is not given. Based on the number of known near neighbours,
only \SDSSb\ is a possible cluster galaxy.

Hence, of the five E+As detected at 21~cm up to now, one is a cluster
member ( SDSS~J230743.41+152558.4) and four are not. Of those not
detected, two are cluster members and four are not. From this result,
it is clear that it is premature to draw conclusions on the
environment by means of neutral hydrogen observations. In order to do
so, a larger sample is required.

\subsection{Optical emission line strengths}

As E+A galaxies are defined by means of their optical spectra, and
more precisely by the equivalent widths of primarily the emission
lines \OII and H$\delta$ \citep{zab96,goto03}, one can investigate any
trend between the neutral hydrogen content and the equivalents
widths. For our sample we list these values in
Table~\ref{EAproperties}. We restate that our E+A galaxies are
compiled from two different samples with each different constraints
concerning the equivalent widths of \OII\ and H$\delta$. In
Table~\ref{othersample} we list the EW(\OII) and EW(H$\delta$) for the
undetected galaxies from \citet{chang01}.

Both groups of detected and undetected E+A galaxies at 21~cm contain a
mixture of \OII\ absorption and emission lines. Similarly, E+A galaxies
with low and high EW(H$\delta$) are detected. We conclude that again a
larger sample is needed in order to investigate any trend between the
optical emission lines and their neutral hydrogen content.

\begin{table}
\begin{center}
\caption{Properties of the E+A galaxies of
\citet{chang01}.}\label{othersample}
\begin{tabular}{lccccccccc}
\hline Galaxy & {\OII}\tablenotemark{a} & H$\delta$\tablenotemark{a} &
H\,{\sc i} mass\tablenotemark{b}\\ & ({\AA}) & ({\AA}) & ($10^9$\, \Msun)\\ \hline
EA01 & 1.80 & 8.98 & 7.1\\
EA02 & 1.25 & 7.98 & $<3.1$\\
EA03 & -0.29 & 8.13 & $<3.9$\\
EA04 & 1.37 & 9.82 & $<2.0$\\
EA11 & 2.16 & 6.96 & $<4.7$\\
\hline
\hline
\end{tabular}
\tablenotetext{a}{{\OII} and H$\delta$ equivalent widths ($\langle
{\rm H}\beta\gamma\delta\rangle$); negative values for {\OII} indicate
absorption}\tablenotetext{b}{The distance $D$, required for
calculating the H{\sc i} mass, is estimated as the Hubble distance, $D
= v_{\rm hel}/H_0$, with $H_0 = 70$ \kms\,Mpc$^{-1}$.}
\end{center}
\end{table}

\subsection{Star-formation rate}

We estimate the star-formation rate (SFR) associated with the observed
H{\sc i} masses using the relation
\begin{equation}
\Sigma_{\rm SFR} \approx 2.5^{\pm 0.7} \times 10^{-10} \left( \frac{\Sigma_{\rm
gas}}{M_\odot \, {\rm pc}^{-2}} \right)^{1.40^{\pm 0.15}} M_\odot\,{\rm
pc}^{-2}\,{\rm yr}^{-1}, \label{sfr}
\end{equation}
\citep{k98}. We substituted the total H{\sc i} mass divided by $\pi
R_{\rm e}^2$, with $R_{\rm e}$ the half-light radius, for the gas
surface density $\Sigma_{\rm gas}$. From the archival F625W HST/ACS
images of EA17, EA18 and EA19, we derived surface brightness profiles
as a function of radius by integrating the light in circular apertures
centered on the galaxy. A similar method was applied to derive the
half-light radii for the SDSS galaxies from Sloan r-band images.  We
fitted seeing or PSF convolved S\'ersic profiles to the surface
brightness profiles of all galaxies. The seeing and PSF profiles are
determined from about 10 stars in each image. We used the surface
brightness profile of a galaxy, extrapolated beyond the last data
point by the best fitting S\'ersic profile, to measure the half-light
radii of these E+As. The SFR surface density, $\Sigma_{\rm SFR}$, was
then converted into a global SFR by multiplying with $\pi R_{\rm
e}^2$. In all cases but SDSS~J230743.41+152558.4 and EA19, we found
the E+As detected at 21~cm to have SFRs in the range
$5-10\,M_\odot\,{\rm yr}^{-1}$ (see Table~\ref{EAproperties}), which
is higher than expected for post-starburst galaxies. Hence, these
gas-rich E+As could be actively forming stars at quite high rates but
the star-formation sites are obscured by dust. Alternatively, although
much gas is present, almost no stars are being formed. The radio
continuum observations of \cite{mo01} rule out star formation at a
rate higher than $1.0-1.5\,M_\odot\,{\rm yr}^{-1}$ in the case of EA17
and EA18. These authors measure a star formation rate of
$1.5\,M_\odot\,{\rm yr}^{-1}$ in EA19. While the star formation rate
of EA19 agrees roughly with our upper limit, the high star formation
rates of EA17 and EA18 derived from their H{\sc i} content are in
clear contradiction with the very low rates derived from radio
continuum observations. Unfortunately, no radio continuum observations
have been performed of {\SDSSa}, {\SDSSb}, and {\SDSSc} so far.

\section{Discussion}\label{discussion}

Three out of the six galaxies observed by us contain detectable
amounts of neutral gas and one is a border case. Both the SDSS
and the Zabludoff samples satisfy very strict selection criteria and
surely do not show any optical evidence for star formation:~these
appear to be true post-starburst galaxies. They do, however, contain
significant amounts of neutral gas. EA19 is not detected by us in
21~cm line emission, with a 3$\sigma$ upperlimit on the \HI\ mass of
$1.2 \times 10^9\,M_\odot$. EA18 was detected by us, with a 21~cm flux
consistent with an \HI\ mass of $2.3 \pm 0.3 \times 10^9\,M_\odot$ of
\HI.

\begin{figure}
\includegraphics[width=8.5cm]{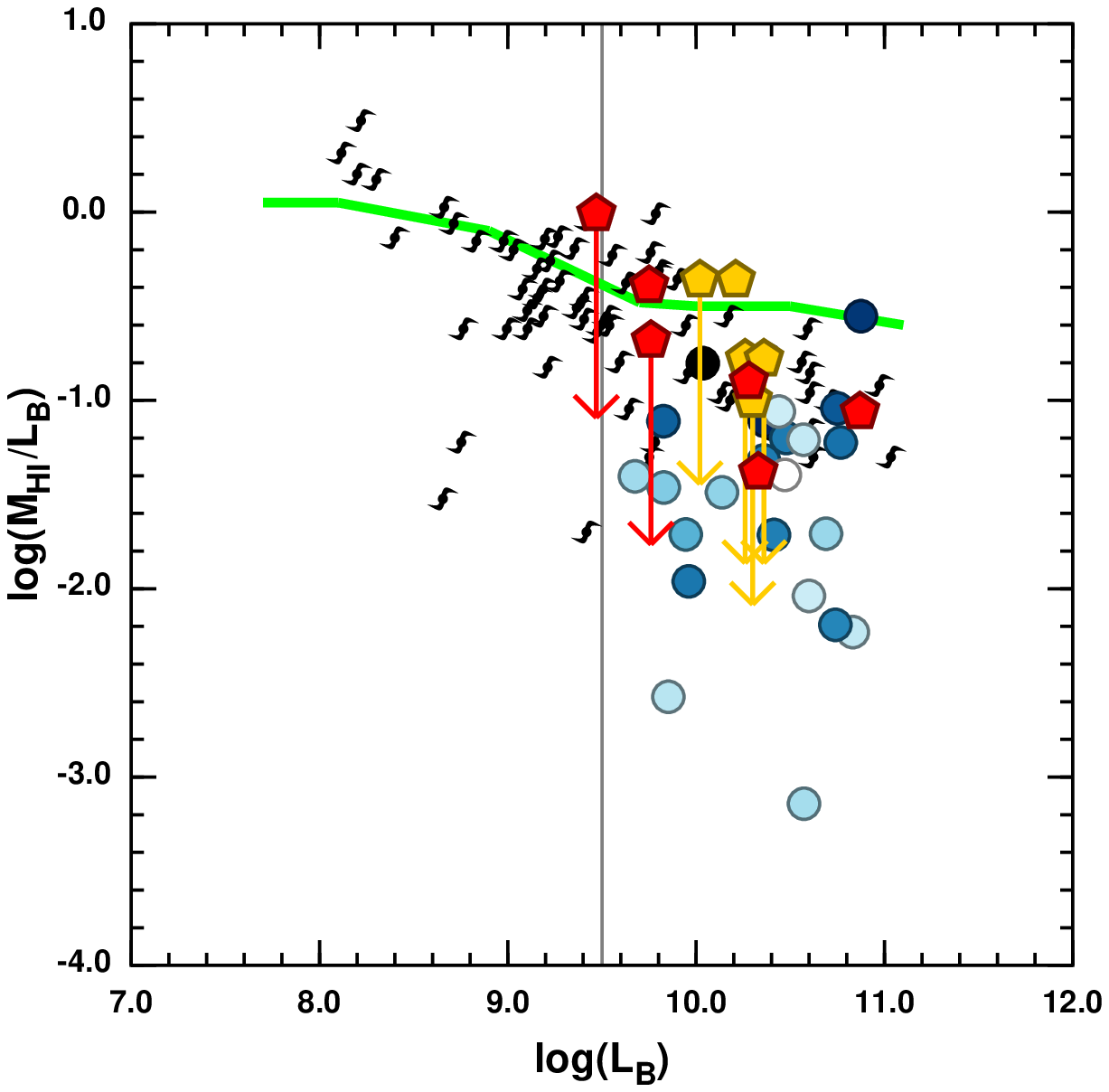}
\caption{$\log(M_{\rm HI}/L_B)$ versus $\log(L_B)$. Spiral galaxies
\citep{h04} are depicted by spiral symbols, elliptical galaxies
\citep{g01} by blue circles (the shade of blue is an indication of
spectroscopic age and the vertical black line shows the luminosity
limit of log($L_B$)=9.5 that was used by \citet{g01}), and E+A
galaxies by red circles (this work) and orange circles
\citep{chang01}. The predicted trend for late-type galaxies by the
semi-analytical models of \citet{ny04} is plotted in
green. \label{logLM}}
\end{figure}

In Fig. \ref{logLM}, we plotted $\log(M_{\rm HI}/L_B)$ versus
$\log(L_B)$, $L_B$ the B-band luminosity expressed in solar B-band
luminosities, for a sample of spiral galaxies \citep{h04}, elliptical
galaxies \citep{g01}, and E+A galaxies \citep[this work
and][]{chang01} along with the predicted $\log(M_{\rm HI}/L_B)$ versus
$\log(L_B)$ relation for late-type galaxies \citep{ny04}. For the E+As
from the SDSS sample, we converted the SDSS g and r magnitudes to the
B-band magnitude using the conversion formulae of \cite{j05}. The
b$_J$ magnitudes of the LCRS E+As were converted into B-band
magnitudes using the relation $m_B = m_J + 0.28 ({\rm B}-{\rm V})$
from \cite{mes90}. Using the mean B$-$V colour of the SDSS E+As,
$\langle ({\rm B}-{\rm V}) \rangle = 0.78$~mag, in this equation, we
find $m_B \approx m_J + 0.22$~mag. The absolute B-band magnitudes
found by applying these conversion formulae are also listed in Table
1. Ellipticals with young spectroscopic ages ($\lesssim 1-2$~Gyr) are
generally more gas-rich than older elliptical galaxies, which
\citet{g01} interpret as evidence for a merger origin for elliptical
galaxies:~the starburst following the merger rapidly consumes the gas
reservoir and subsequent quiescent star formation consumes the gas at
a much slower rate. It should be noted that \citet{g01} include only
ellipticals brighter than $M_B = -18.5$~mag, or $\log(L_B) = 9.5$, in
their sample because fainter galaxies are likely to have experienced
different evolutionary histories. The resulting sample of ellipticals
spans about the same luminosity range as the E+A samples.

As is clear from Fig. \ref{logLM}, some E+As are more gas-rich than
most young elliptical galaxies. This is most likely not a selection
effect, since the E+A data set was assembled based on optical,
spectral properties and the elliptical galaxies were selected
according to optical morphological considerations, not on H{\sc i}
mass. One would, moreover, expect the \citet{g01} data set to be
complete for galaxies with large H{\sc i} masses. We performed a
Kolmogorov-Smirnov (KS) and Wilcoxon (W) test on the $\log(M_{\rm
HI}/L_B)$ distribution (within the same magnitude limits, both in
$L_B$ and $\log(M_{\rm HI}/L_B)$, i.e. in the region defined by the
detected E+As) and found that the distribution of E+A galaxies (i)
corresponds with that of spiral and young elliptical galaxies with a
significance of respectively 99\% (KS) or 94\% (W) and 65\% (KS) or
79\% (W) and (ii) differs from that of old elliptical galaxies with a
confidence level of 79\% (KS) or 78\% (W). This statistical test and
Fig. \ref{logLM} might suggest a gas depletion time sequence, with
E+As being observed less than $\sim 500$~Myr after the termination of
the starburst that followed the putative merger \citep{yang04}, young
ellipticals after $\lesssim 1-2$~Gyr, and old ellipticals at later
times. However, one should note that the difference between the
distributions of E+As and elliptical galaxies has just a significance
of 1$\sigma$. In order to derive conclusive evidence for this
suggestion, a larger statistical sample of H{\sc i} masses and
spectroscopic ages of E+As is essential. Nevertheless, if correct, this
ties together with the morphological study of 5 E+As from the
Zabludoff (1996) sample with HST by \citet{yang04}, who suggest that
E+As are likely to evolve to elliptical galaxies with power-law
density profiles.

One should also note that the end of a starburst does not necessarily
require the complete exhaustion of the neutral gas reservoir. This
could be explained by the fact that neutral gas itself is not the raw
material for star formation:~stars form in the dense cores of
molecular clouds. \citet{ko02} have obtained CO($1\!\rightarrow\! 0$)
and HCN($1\!\rightarrow\! 0$) observations of the nearby
post-starburst galaxy NGC5195, which, interestingly, forms an
interacting pair with NGC5194. These authors note a central decrease
of the mass fraction of dense molecular cores, leaving only diffuse
molecular gas, evidenced by a very low central $L_{\rm HCN}/L_{\rm
CO}$ value. Most likely, an intense central burst of star formation
$\sim 1$~Gyr ago, responsible for the observed population of A stars,
evaporated the dense molecular clouds, which are the sites where
massive stars form. This effectively stopped further star formation,
although large amounts of diffuse neutral and molecular gas remain. A
similar mechanism may be responsible for switching off the starburst
in E+A galaxies without the necessity of consuming the complete gas
reservoir. The remaining \HI\ reservoir in some E+As may eventually
lead, after gas has been allowed to condense into molecular cores, to
further episodes of star formation. Eq. (\ref{sfr}), which is based on
observations of normal spiral galaxies that host a balanced mix of
neutral and molecular gas, may therefore not be applicable to
post-starburst galaxies.

\section{Conclusions}\label{concl}

We present deep single-dish \HI\ observations of a sample of six
nearby E+A galaxies ($0.05<z<0.1$). We find {\HI} masses of $6.5 \pm
0.8 \times 10^9\,M_\odot$ for {\SDSSa}, $0.9 \pm 0.3 \times
10^9\,M_\odot$ for {\SDSSb}, $2.7 \pm 0.3 \times 10^9\,M_\odot$ for
{\SDSSc}, and $2.3 \pm 0.3 \times 10^9\,M_\odot$ for EA18. The
$3\sigma$ upperlimits for the gas content of EA17 and EA19 are $2.9
\times 10^9\,M_\odot$ and $1.2 \times 10^9\,M_\odot$ respectively,
assuming a total velocity width of 450 {\kms} (the average velocity
width of the detected galaxies). The three galaxies from the SDSS
sample satisfy very strict selection criteria and, for all practical
purposes, can be considered to be truly post-starburst galaxies.
 
The E+A galaxies detected in 21~cm line emission are almost as
gas-rich as spiral galaxies with comparable luminosities. By plotting
these E+As, spiral galaxies, and elliptical galaxies in a $\log(M_{\rm
HI}/L_B)$ versus $\log(L_B)$ diagram, and performing a
Kolmogorov-Smirnov and Wilcoxon test we suggest that existence of a
gas depletion time sequence, with E+As being observed very shortly
after the termination of the starburst that ensued from the putative
merger, young ellipticals after $\lesssim 1-2$~Gyr, and old
ellipticals at even later times. This would tie together with the
morphological study of 5 E+As from the Zabludoff (1996) sample with
HST by \citet{yang04}, who suggest that E+As are likely to evolve to
elliptical galaxies with power-law density profiles. However, the
conclusions drawn here are just based on 1$\sigma$ trends. A larger
sample of \HI\ detections and spectroscopic ages of E+As is required
to investigate this hypothesis.

The presence of \HI\ in an E+A galaxy can be explained in two ways.\\
(i) We can interpret the lack of on-going star-formation in E+A
galaxies, suggested by previous radio continuum observations
\citep{mo01} and indirectly by their selection criteria, and the fact
that the end of the starburst does not necessarily require the
complete exhaustion of the neutral gas reservoir, as being due to the
effect the starburst has on the dense molecular cores which are
responsible for the massive star formation. An intense burst of star
formation can evaporate the dense molecular clouds, effectively
stopping further star formation, even though copious amounts of
diffuse neutral and molecular gas remain. The remaining \HI\ reservoir
in some E+As may eventually lead, after the gas has again condensed
into molecular cores, to further episodes of star formation. This may
indicate that E+As are observed in the inactive phase of the
star-formation duty cycle.\\(ii) A second possibility is what
previously has been proposed by Couch \& Sharples (1987) and Blake et
al. (2004). There might still be on-going star-formation associated
with the presence of {\HI} gas, which is hinted from the high SFR (see
Table~\ref{EAproperties}) of our E+A galaxies; however it cannot be
observed since it is obscured by dust \citep{smg99}, which is
suggested by the ACS images in Fig.~\ref{ima_EA}, and/or optical
emission lines might be a poor way of isolating true post-starbust
systems. In this case E+A galaxies are in the active star-formation
phase and will presumably exhaust all their gas content.

\acknowledgements 

PB acknowledges financial support from the Bijzonder OnderzoeksFonds
(BOF). DM is supported by the MAGPOP EU Marie Curie Training and
Research Network. SD acknowledges financial support from the Fonds
voor Wetenschappelijk Onderzoek -- Vlaanderen (FWO). This research was
performed while D. J. P. held a National Research Council Research
Associateship Award at the Naval Research Laboratory. Basic research
in astronomy at the Naval Research Laboratory is funded by the Office
of Naval Research. This work has made use of the NASA/IPAC
Extragalactic Database (NED) which is operated by the Jet Propulsion
Laboratory, California Institute of Technology, under contract with
the National Aeronautics and Space Administration. Based on
observations made with the NASA/ESA Hubble Space Telescope, obtained
from the data archive at the Space Telescope Institute. STScI is
operated by the association of Universities for Research in Astronomy,
Inc. under the NASA contract NAS 5-26555.

\end{document}